\documentclass[11pt, a4paper]{article}
\usepackage{amssymb}
\usepackage{eurosym}
\usepackage{amsfonts}
\usepackage{amsmath}
\usepackage{array}
\usepackage{calc}
\usepackage{array}
\usepackage{indentfirst}
\usepackage{graphicx}
\usepackage{hyperref}

\usepackage{authblk}
\author[1]{Remigiusz Durka \thanks{remigiusz.durka@ift.uni.wroc.pl}}
\author[1,2]{Jerzy Kowalski-Glikman \thanks{jerzy.kowalski-glikman@ift.uni.wroc.pl}}
\affil[1]{Institute for Theoretical Physics, University of Wroc\l{}aw, pl.\ M.\ Borna 9, 50-204 Wroc\l{}aw, Poland}
\affil[2]{National Centre for Nuclear Research, Pateura 7, 02-093 Warsaw, Poland}
\title{Resonant algebras in Chern-Simons model of topological insulators}
\date{\today}

\begin{document}
\maketitle
\begin{abstract}
	This paper explores the possibility of using Maxwell algebra and its generalizations called resonant algebras for the unified description of topological insulators. We offer the natural action construction, which includes the relativistic Wen-Zee and other terms, with adjustable coupling constants. By gauging all available resonant algebras formed by Lorentz, translational and Maxwell generators $\{J_a, P_a, Z_a\}$ we present six Chern-Simons Lagrangians with various terms content accounting for different aspects of the topological insulators.
	
	Additionally, we provide complementary actions for another invariant metric form, which might turn out useful in some generalized (2+1) gravity models.
\end{abstract}

\section{Introduction}
Chern-Simons (CS) model plays an important role in the theory of Quantum Hall systems \cite{Tong:2016kpv}. As an effective field theory, it is capable of describing various physical effects appearing in the intriguing materials called topological insulators \cite{Hasan:2010xy, Abanov:2014ula, Gromov:2015fda, Blasi:2011pf, Salgado-Rebolledo:2019kft}, which behave like insulators in the bulk but render non-trivial conducting on the edges. In the topological insulator action, the Hall current is described by the CS$[A]$ term, where $A_\mu$ is the electromagnetic potential. One can also introduce the torsional term, accounting for Hall fluid viscosity \cite{Hughes:2012vg,Geracie:2014mta, Cappelli:2015ocj}, and a term of the form of CS$[\widehat{\omega}]$ called "gravitational" by the Hall effect community, where $\widehat{\omega}$ is another O(2) connection. In addition, we include the so-called Wen-Zee term \cite{Wen:1992ej} encoding direct interaction between $A$ and $\widehat{\omega}$.

The Chern-Simons action configuration is fixed when one chooses the relevant gauge symmetry. The problem is that the usual "gravitational"($\omega$) and  "electromagnetic" ($A$) actions do not offer a natural interaction between these two different theories. In the literature, such interaction is included by hand in the form of the so-called Wen-Zee term \cite{Wen:1992ej}
\begin{equation}\label{0}
 S_{\mathrm{WZ}}=\frac{\nu \bar{s}}{4\pi }\int d^{3}x\,\epsilon ^{\mu \rho \sigma
}A_{\mu }\partial _{\rho }\widehat{\omega}_{\sigma }
\end{equation}
where $\widehat{\omega}_{\sigma }=\frac{1}{2}\sigma_{\hat{a}\hat{b}}\,\omega^{\hat{a}\hat{b}}_\sigma$ represents O(2) spin connection transforming as a $U(1)$ gauge field, while $A_\mu$ is electromagnetic gauge field with $\nu$ being filling factor and $\bar{s}$ describing average orbital spin coming from the angular momentum contribution due to the cyclotron orbits in the Hall conductance. Note that presence of \eqref{0} in the action is merely due to some heuristic motivation and not the result of direct CS construction.

In the recent paper \cite{Palumbo:2016nku} G. Palumbo delivers an interesting take on that subject, exploring the application of the so-called Maxwell gauge symmetry to the  topological insulators. The basic idea is to embed the two abelian gauge groups into a larger gauge group in order to obtain the description of Quantum Hall system in terms of a single Chern-Simons theory. Unfortunately, in the paper \cite{Palumbo:2016nku} this goal is only partially achieved. The aim of this letter to complete this construction.

To incorporate the Wen-Zee term into the unified framework it has been proposed in \cite{Palumbo:2016nku} to use the Chern-Simons theory of the so-called the Maxwell algebra. This algebra was introduced in the 1970's \cite{Bacry:1970ye, Schrader:1972zd} and is defined by  generators $\{J_{ab},P_{a},Z_{ab}\}$. It was thoroughly analyzed in a wide range of contexts (see \cite{Salgado:2014qqa,Diaz:2012zza,Edelstein:2006se,deAzcarraga:2010sw,Kamimura:2011mq,deAzcarraga:2012zv, Concha:2014vka, Concha:2018zeb, Aviles:2018jzw}, along with many further generalizations \cite{Durka:2011nf,Durka:2011va,Durka:2012wd,Concha:2016hbt,Concha:2016kdz,Concha:2018jjj}). Its novelty concerns the  commutator of two translations, which takes the form
\begin{equation}\label{01}
  [P_{a},P_{b}]=Z_{ab}\,,
\end{equation}
where $Z_{ab}$ are additional Maxwell algebra generators satisfying
\begin{equation}\label{02}
  [Z_{ab},P_{c}]=0\,,\quad [Z_{ab},Z_{cd}]=0
\end{equation}
(Notice the difference between Maxwell algebra and the (A)dS one, where the right hand side of \eqref{01} is proportional to the Lorentz generator $J_{ab}$ and it vanishes for the Poincar\'e algebra.)

In \cite{Palumbo:2016nku}, the electromagnetic gauge field $A_{\mu }$ has to be immersed in the $\frac{1}{2}%
\hat{A}_{\mu }^{ab}Z_{ab}$, part of the full connection describing the new non-abelian \textit{Maxwel algebra}-valued field. This immersion made it possible to reproduce the Wen-Zee term, moreover, in the relativistic form. In such a formulation of the topological insulator model, the Maxwell
algebra provided essential structure and field content, assuring the interaction term between the geometry and gauge field. Unfortunately, at the same time, it failed in reproducing directly the gauge field contribution,
\begin{equation}\label{03}
\frac{\nu}{4\pi }\int d^{3}x\,\epsilon^{\mu \rho \sigma}A_{\mu }\partial _{\rho }A_{\sigma }
\end{equation}
since the vanishing of the corresponding invariant tensor $\langle Z_{a},Z_{b}\rangle =0$ does not allow for such term. As this scenario corresponds to the trivial topological insulator, to resolve this apparent obstacle, the final action in \cite{Palumbo:2016nku} was composed as a sum of two Chern-Simons theories. 

 In this paper, we offer a generalization of this construction capable of resolving some of the issues. We argue, that to obtain a more suitable description, one needs to go through a standard (single) Chern-Simons formulation. The only difference lies in exploiting other algebra of a similar type. Such algebra was introduced by Soroka and Soroka \cite{Soroka:2004fj, Soroka:2006aj} and eventually will represent the case with the most general term content.

Moreover, the semigroup expansion framework \cite{Izaurieta:2006zz} showed that one can have altogether 6 algebras \cite{Durka:2016eun} resulting from adding the new generator $Z_{ab}$ to $J_{ab}$ and $P_{a}$. These different algebras will lead to the various configuration of the existing terms in the Lagrangians, as the particular algebras give different invariant tensors.

\section{Gauging the resonant algebras}

We start with the extended definition of the gauge connection
\begin{align}
\mathcal{A}=\omega^{a}J_{a}+\frac{1}{\ell }\,e^{a}P_{a}+\hat{A}^{a}Z_{a}\,,\label{connection}
\end{align}
where gauge algebra indices are given by $a=0,1,2$. The resonant algebra in 2+1 dimensions is defined by three groups of generators.  $J_{a}$ are generators of  Lorentz transformations (rotations and boosts), $P_{a}$ are the generators of spacetime translations, while $Z_{a}$ are the new Maxwell
generators. Such definition naturally requires an introduction of the length parameter $\ell $ to make the dimensions right for the dreibein term. Since we are dealing with the $(2+1)D$ theory, we can consider the dual generators $J_{a}=\frac{1}{2}\epsilon _{abc}J^{bc}$ as well as the redefinition of the spin connection $\omega _{a}=\frac{1}{2}\epsilon _{abc}\omega ^{bc}$ and, analogously, $Z_{a}=\frac{1}{2}\epsilon _{abc}Z^{bc}$ and the gauge field $\hat{A}_{a}=\frac{1}{2}\epsilon_{abc}\hat{A}^{bc}$.

For a given set of generators $X_A=\{J,P,Z\}$ the commutation relations can be written as $[X^A_a,X^B_b]=\left(f^{AB\;}_{\;\;\;\;C} X^C_c\right)\epsilon_{ab\,}^{\;\;\;\;c}  $. Depending on the right hand side of the $[P_a,P_b]$ commutator we can organize algebras according to the resulting sub-algebras of $\{J_a,P_a\}$ (see details concerning derivation and the notation in \cite{Durka:2016eun}). Gathering schematically the outcomes $f^{AB\;}_{\;\;\;\;C} X^C$ in the table
\begin{equation}
\begin{tabular}{c|ccc}
$[. \,, . ]$ & J & P & Z \\ \hline
J & . & . & . \\
P & . & . & . \\
Z & . & . & .
\end{tabular}
\end{equation}
gives us 
%$[X^A_a,X^B_b]=f^{AB\;}_{\;\;\;\;C} X^C_c \epsilon_{ab\,}^{\;\;\;\;c}$
\begin{itemize}
	\item $4\times $ Poincar\'{e}-like algebras (i.e. having $[P_a,P_b]=0$), which we denote as type: $B_{4}$, $%
	\tilde{B}_{4}$, $\tilde{C}_{4}$, and $C_{4}\equiv Poincare\oplus Lorentz$
	\begin{equation}
	\begin{tabular}{c|ccc}
	\framebox{$B_{4}$} & J & P & Z \\ \hline
	J & J & P & Z \\
	P & P & 0 & 0 \\
	Z & Z & 0 & 0
	\end{tabular}
	\quad
	\begin{tabular}{c|ccc}
	\framebox{$\tilde{B}_{4}$}& J & P & Z \\ \hline
	J & J & P & Z \\
	P & P & 0 & P \\
	Z & Z & P & 0
	\end{tabular}
	\quad
	\begin{tabular}{c|ccc}
	\framebox{$\tilde{C}_{4}$}& J & P & Z \\ \hline
	J & J & P & Z \\
	P & P & 0 & 0 \\
	Z & Z & 0 & Z
	\end{tabular}
	\quad
	\begin{tabular}{c|ccc}
	\framebox{$C_{4}$} & J & P & Z \\ \hline
	J & J & P & Z \\
	P & P & 0 & P \\
	Z & Z & P & Z
	\end{tabular}
	\end{equation}
	
	\item no AdS-like algebra, which would contain $[P_a,P_b]=\epsilon_{ab\,}^{\;\;\;\;c}J_{c}$
	
	\item $2\times$ Maxwell-like algebras (i.e. containing $[P_a,P_b]=\epsilon_{ab\,}^{\;\;\;\;c}Z_{c}$): of type $\mathfrak{B}_{4}$ (original
	Maxwell algebra introduced in the 70's) and type $\mathfrak{C}_{4}\equiv
	AdS\oplus Lorentz$ (introduced by Soroka-Soroka and shown to represent under change of basis the direct sum of two algebras)
\end{itemize}
\begin{equation}
\begin{tabular}{c|ccc}
\framebox{$\mathfrak{B}_{4}$}& J & P & Z \\ \hline
J & J & P & Z \\
P & P & Z & 0 \\
Z & Z & 0 & 0
\end{tabular}
\qquad
\begin{tabular}{c|ccc}
\framebox{$\mathfrak{C}_{4}$}& J & P & Z \\ \hline
J & J & P & Z \\
P & P & Z & P \\
Z & Z & P & Z
\end{tabular}
\end{equation}
 We thus have in total 6 algebras, of which we are now going to focus solely on the last one, as it offers maximal content of terms, postponing commenting about other possibilities to the end of this section. The algebra $\mathfrak{C}_{4}$ is defined explicitly in (2+1)-dimensions by the following commutators
\begin{align}
\hspace{0.2cm}[J_{a},J_{b}]& =\epsilon _{abc}J^{c},\hspace{0.6cm}%
[J_{a},P_{b}]=\epsilon _{abc}P^{c}\, , \\
\lbrack P_{a},P_{b}]& =\epsilon _{abc}Z^{c},\hspace{0.6cm}%
[J_{a},Z_{b}]=\epsilon _{abc}Z^{c}\, ,\\
\hspace{0.2cm}[P_{a},Z_{b}]& =\epsilon _{abc}P^{c},\hspace{0.6cm}%
[Z_{a},Z_{b}]=\epsilon _{abc}Z^{c}\, ,
\end{align}
where $a,b,c=0,1,2$. This algebra differs from the Maxwell one in the last two relations, i.e. $[P_{a},Z_{b}]$ and $%
[Z_{a},Z_{b}]$, which are zero. Note that the four other cases of the "Poincar\'{e}-like" algebras share $%
[P_{a},P_{b}]=0$, but they have other outcomes of commutators $P_a$ with $Z_b$ and $Z_a$ with $Z_b$.

The (internal) invariant metric for the Soroka-Soroka $\mathfrak{C}_{4}$ algebra, given by a bilinear form $\Omega _{AB}=\langle X_{A},X_{B}\rangle $, is identified through the following relations:
\begin{align}
\langle J_{a}\,J_{b}\rangle & =\alpha _{0}\,\eta _{ab}\, ,\hspace{%
0.6cm}\langle P_{a}\,P_{b}\rangle =\alpha _{2}\,\eta _{ab}\, ,  \notag \\
\langle J_{a}\,Z_{b}\rangle & =\alpha _{2}\,\eta _{ab}\, ,\hspace{0.6cm}\langle
Z_{a}\,Z_{b}\rangle =\alpha _{2}\,\eta _{ab}\, ,
\end{align}
where $\alpha _{0}$ and $\alpha _{2}$ are the arbitrary real parameters.
Particular set of $\alpha $ constants can be seen as the result of the
semigroup expansion procedure applied to the original (A)dS invariant metric (see \cite{Salgado:2014qqa,Diaz:2012zza,Concha:2018jjj,Durka:2016eun}),
or just by the realization of $\langle \lbrack X_{A},X_{B}],X_{C}\rangle
=\langle X_{A},[X_{B},X_{C}]\rangle $.

In general, there is yet another type of the invariant metric, namely
\begin{align}
\langle J_{a}\,P_{b}\rangle =\alpha _{1}\,\eta _{ab}\, ,\qquad\mbox{and}\qquad\langle
Z_{a}\,P_{b}\rangle =\alpha _{1}\,\eta _{ab}\, ,
\end{align}
which leads to the appearance of terms like $R_{a}e^{a}=\frac{1}{2}R^{ab}e^{c}\epsilon
_{abc}$. We set it to zero by the means of fixing $\alpha _{1}=0$, as they are are irrelevant to the subject of the topological insulators \cite{Palumbo:2016nku}. However, for the completeness, we will include separately a complementary action construction based on these invariants too, which might find use in the generalized $(2+1)$-gravity models.

The gauge connection $\mathcal{A_{\mu }}$ defined in Eq.~\eqref{connection} takes values in a given algebra (which we specified here to be $\mathfrak{C}_{4}$), whereas the field strength tensor $\mathcal{F_{\mu \nu }}$ is given by
\begin{align}
\mathcal{F_{\mu \nu }}=\mathcal{F_{\mu \nu }}^{A}X_{A}=R_{\mu \nu }^{a}J_{a}+%
\frac{1}{\ell }T_{\mu \nu }^{a}P_{a}+H_{\mu \nu }^{a}Z_{a}\,.
\end{align}
We denote the standard Riemann two-form as $R^{a}=d\omega^a+\frac{1}{2}\epsilon^{abc}\omega_b\wedge \omega_c$, while particular $H^{a}$ form will depend on the specific algebra. For the Soroka-Soroka algebra $H^{a}$ will be expressed as
\begin{align}
H^a=D_{\omega }\hat{A}^{a}+\frac{1}{2}\epsilon^a_{~bc}\,\hat{A}^{b}\wedge \hat{A}^{c} +\frac{1}{2\ell ^{2}}\epsilon^a_{~bc}\,e^b\wedge e^c \,,
\end{align}
whereas for the Maxwell algebra it will be reduced to $H^a=D_{\omega}\hat{A}^{a}=d\hat{A}^{a}+\epsilon^{a}_{~bc}\,\omega^b \wedge \hat{A}^{c}$. Depending on the algebra, also notion of the \textit{torsion} two-form $T^{a}$ can be altered. Every time the algebra delivers $[Z_{a},P_{b}]=\epsilon _{abc}P^{c}$ (as it happens for $\mathfrak{C}_4$) it will result in the extra $\epsilon^{a}_{~bc}\,\hat{A}^b \wedge e^{c}$ term added to the standard part of $de^{a}+\epsilon^{a}_{~bc}\,\omega^b \wedge e^{c}$.

We choose our Lagrangian to be that of Chern-Simons model \cite{Chamseddine:1990gk, Zanelli:2012px, Witten:1988hc}
\begin{align}
\mathcal{L}_{CS}^{2+1}=\frac{k}{4\pi}\langle \mathcal{A}\wedge d\mathcal{A}+\frac{1}{3}\mathcal{A%
}\wedge [\mathcal{A},\mathcal{A}]\rangle \,,
\end{align}
where $\langle \ldots \rangle $ denotes the particular invariant tensor. Note that, the Palumbo's final action \cite{Palumbo:2016nku} for the Maxwell algebra was written as:
\begin{align}
S_{Palumbo}[e,\omega ,\hat{A}]& =\frac{k}{4\pi}\int \mathrm{tr[\varrho
	_{1}CS(\omega )+\varrho _{2}\,e \wedge D_{\omega }e}+\varrho _{3}\widehat{A}\wedge (R+\varrho _{4}\,e \wedge e)+\varrho _{5}%
\widehat{A}\wedge D_{\omega }\widehat{A}]\,.
\end{align}
This was, however, achieved by including the additional inverse of bilinear form $\Omega^{-1}_{AB}$ (which can not be used as the internal metric) along with some further complications regarding the constants. The proper action for the Maxwell $\mathfrak{B}_4$ algebra can be found in \cite{Concha:2018zeb, Aviles:2018jzw}. 

In turn, we find that (up to the boundary terms) the Chern-Simons action gauged straightforwardly using the Soroka-Soroka algebra \cite{Diaz:2012zza,Concha:2018jxx, Concha:2018jjj} is of the form
\begin{align}
S_{\mathfrak{C}_4}[e,\omega ,\hat{A}]& =\frac{k}{4\pi}\int \alpha_0\,CS(\omega )\nonumber\\
	&+\frac{k}{4\pi} \int \alpha_2\,\left( \frac{1}{\ell^2}\,e_a \wedge D_{\omega }e^a+2\hat{A}_a \wedge (R^a+\frac{1}{2\ell^2}\epsilon^{abc}\,e_b\wedge e_c)+ \hat{A}_a\wedge D_{\omega }\hat{A}^a\right)\nonumber\\
	&+\frac{k}{4\pi} \int \alpha_2\, \frac{1}{3}\,\epsilon^{abc}\,\hat{A}_a \wedge\hat{A}_b\wedge\hat{A}_c\,.\label{CS-SS}
\end{align}
Interestingly, the set of the resulting constants is reduced just to $k, \alpha_0,\alpha_2$ and $\ell$. The variations over fields provide the field equations \begin{align}
\delta e_a:\qquad 0&=\frac{2}{\ell^2}\alpha_2 \left(D_{\omega }e^a+\epsilon^{abc}\,\hat{A}_b\wedge e_c
\right)\\
\delta \hat{A}_a:\qquad 0&=2\alpha_2 \left(R^a+(D_{\omega }\hat{A}^a+\frac{1}{2}\epsilon^{abc}\,\hat{A}_b\wedge \hat{A}_c+\frac{1}{2\ell^2}\epsilon^{abc}\,e_b\wedge e_c)\right)\\
\delta \omega_a:\qquad 0&=2\alpha_0\, R^a+2\alpha_2\,\left(D_{\omega }\hat{A}^a+\frac{1}{2}\epsilon^{abc}\,\hat{A}_b\wedge \hat{A}_c+\frac{1}{2\ell^2}\epsilon^{abc}\,e_b\wedge e_c\right)\,.
\end{align}

In action above, one finds the standard torsional Chern-Simons term $e_a \wedge D_\omega e^a$ along with the Chern-Simons term for the $\omega$
\begin{align}
CS(\omega)=\omega^a\wedge d \omega_a+\frac{1}{3}\,\epsilon_{abc}\,\omega^a\wedge \omega^b\wedge\omega^c\,.
\end{align}
These two terms in the gravity context form the "exotic" Chern-Simons action \cite{Zanelli:2012px}, which might be added to the standard 3D gravity $CS(\omega,e)=\frac{1}{2}R^{ab} \wedge e^{c}\epsilon
_{abc}=R_{a}\wedge e^{a}$ \cite{Mielke:1991nn}. Here they will  play a leading role in describing Hall viscosity.

Having the $\mathfrak{C}_4$ action \eqref{CS-SS} we can now relate it to the topological insulators. Before doing so, let us first gather all other resonant algebras cases and summarize them in a single table (with "-" meaning lack of a particular term)
\begin{table}[h]
	\caption{Lagrangian content for the $\{J_a, P_a, Z_a\}$ resonant algebras}
\begin{center}
	\begin{tabular}{l|c|c|c|c|c|c}
	& ~~~~$B_4$~~~~ & ~~~~$\tilde{B}_{4}$~~~~ & ~~~~$\tilde{C}_{4}$~~~~ & ~~~~$C_4$~~~~ & ~~~~$\mathfrak{B}_4$~~~~ & ~~~~$\mathfrak{C}_4$~~~~ \\ \hline
	$C(\omega)$ & $\alpha_0$ & $\alpha_0$ & $\alpha_0$ & $\alpha_0$ & $\alpha_0$          & $\alpha_0$   \\
	$2\,\hat{A}_a R^a$   & $\alpha_2$ & $\alpha_2$ & $\alpha_2$ & $\alpha_2$ & $\alpha_2$          & $\alpha_2$          \\
	$\hat{A}_a D_\omega \hat{A}^a$  & -          & - & $\alpha_2$ & $\alpha_2$          & -                   & $\alpha_2$          \\
	$\frac{1}{3}\epsilon^{abc}\hat{A}_a\hat{A}_b\hat{A}_c$  & -          & - & $\alpha_2$ & $\alpha_2$          & -                   & $\alpha_2$ \\
	$\frac{1}{\ell^2}\epsilon^{abc}\hat{A}_a e_b e_c$  & -          & -          & -          & -          & -                   & $\alpha_2$          \\
	$\frac{1}{\ell^2}e_a T^a$   & -          & -          & -          & -          & $\alpha_2$          & $\alpha_2$        \\
\end{tabular}\end{center}
\end{table}\\
Already at this point (still before using an ansatz relating the gauge field $A_\mu$ to the non-abelian $\hat{A}^a_\mu$ field), it is obvious that the original Maxwell algebra, denoted as $\mathfrak{B}_4$, is not able to produce non-trivial description of topological insulators. An algebra with $[Z_a,Z_b]=0$ and $\langle Z_a\,Z_b\rangle  = 0$ simply cannot produce the $A dA$ term. Interestingly, two other algebras, $\tilde{C}_{4}$ and $C_{4}$, are able to deliver required $\langle Z_a\,Z_b\rangle  \neq 0$, but they can not account for the torsional Hall viscosity part carried by the missing $e_a\wedge T^a$ term. Only the use of Soroka-Soroka $\mathfrak{C}_4$ algebra assures the maximal term content and the widest scope of described phenomena.

Similarly, we can supplement a complementary action representing a counterpart with non-vanishing $\langle J_{a}\,P_{b}\rangle =\langle Z_{a}\,P_{b}\rangle =\alpha _{1}\,\eta _{ab}\, $, which might be used for $(2+1)$-gravity models.
\begin{table}[h]
		\caption{Lagrangian counter-content for the $\{J_a, P_a, Z_a\}$ resonant algebras}
	\begin{center}
		\begin{tabular}{l|c|c|c|c|c|c}
			& ~~~~$B_4$~~~~ & ~~~~$\tilde{B}_{4}$~~~~ & ~~~~$\tilde{C}_{4}$~~~~ & ~~~~$C_4$~~~~ & ~~~~$\mathfrak{B}_4$~~~~ & ~~~~$\mathfrak{C}_4$~~~~ \\ \hline
			$\frac{2}{\ell} e_a R^a$ & $\alpha_1$  & $\alpha_1$  & $\alpha_1$  & $\alpha_1$  & $\alpha_1$           & $\alpha_1$    \\
			$\frac{2}{\ell} e_a D_\omega \hat{A}^a$  & -          & $\alpha_1$  & - & $\alpha_1$         & -                   & $\alpha_1$          \\
			$\frac{1}{\ell} e_a\hat{A}_b\hat{A}_c\, \epsilon^{abc}$  & -          & - & - & $\alpha_1$           & -                   & $\alpha_1$  \\
			$\frac{1}{3\ell^3} e_a e_b e_c\, \epsilon^{abc}$  & -          & -          & -          & -          & -                   & $\alpha_1$           \\
	\end{tabular}\end{center}
\end{table}

When it comes for the second table, applied in AdS gravity context, we would set $k=\frac{1}{4G}$ for $\mathcal{L}=\frac{k}{4\pi} CS[e,\omega]$, which obviously would settle down the constants as $\alpha_1=\frac{\ell}{16\pi G}$ and $\Lambda=-\frac{1}{\ell^2}$. Looking at the tables separately, we face similarity of term content between some algebras (for example between $B_4,\tilde{C}_4$ and $\mathfrak{B}_4$ in the first table, as well as between $B_4$ and $\mathfrak{B}_4$ in the second one). This is just apparent because taking both tables together always produces unique action construction. So far only $\mathfrak{B}_4$ and $\mathfrak{C}_4$ configurations had a fair share of thorough analysis in the literature. Other cases might provide a simpler setup in some applications and find use for example in supergravity.

\section{From Chern-Simons model to topological insulators}
	
Obtaining the topological insulators description from the action \eqref{CS-SS} requires additional immersion of the standard $U(1)$ gauge field $A_\mu$ into the Maxwell field $\hat{A}_\mu^a$. Following \cite{Palumbo:2016nku} we do that by choosing a gauge breaking ansatz
\begin{align}
\hat{A}^0_\mu=A_\mu,\quad\quad \hat{A}^1_\mu=0,\qquad\mbox{and}\qquad \hat{A}^2_\mu=0\,.
\end{align}
Such explicit gauge symmetry breaking is far from being satisfactory, and eventually should be replaced by some dynamical mechanism. It seems natural to employ the Higgs mechanism in this context. This mechanism would produce a massless gauge field and massive ones, whose interpretation in the context of the topological insulators is not clear. We leave the detailed discussion of the Higgs mechanism to the future publication.

We are left now with the identification of the coupling constants. In standard case of $S_{C S}[A]=\frac{k}{4 \pi} \int d^{3} x \epsilon^{\mu\rho\sigma } A_{\mu} \partial_{\rho} A_{\sigma}$ one computes the current that arises from the Chern-Simons term $J_{i}=\frac{\delta S_{C S}[A]}{\delta A_{i}}=\frac{k}{2 \pi} \epsilon_{ji} E_{j}$ and the charge density $J_0=\frac{\delta S_{C S}[A]}{\delta A_{0}}=\frac{k}{2 \pi} B$ (see  \cite{Tong:2016kpv}). This means that the Hall conductivity takes the value of $\sigma_{xy}=\frac{k}{2\pi}$. Then the Chern-Simons level corresponds to the filling factor $\nu$ of the Landau levels, accordingly to $k= \frac{e^2\nu}{\hbar}$.

In most of the conventions: $\hbar=e=1$, so in our case the corresponding part of the action
 \begin{align}
\frac{k}{4\pi}\alpha_2\, \int  \hat{A}_a\wedge D_{\omega }\hat{A}^a= \frac{k}{4\pi}\alpha_2\, \int  \hat{A}_0\wedge D_{\omega }\hat{A}^0=\frac{k}{4\pi}\alpha_2\, \int A\wedge d A\,,
\end{align}
after setting $k \alpha_2=\nu $, restores the expected term
\begin{equation}
\frac{\nu }{4\pi}\int A d A\,.
\end{equation}
The same happens for the Wen-Zee term $S_{WZ}=\frac{\nu \bar{s}}{2\pi}\int A\,d\widehat \omega$, now appearing in a generalized form
\begin{align}
\frac{k}{2\pi}\alpha_2\, \int \hat{A}_0 \wedge (R^0+\frac{1}{2\ell^2}\epsilon^{0bc}\,e_b\wedge e_c)=\frac{\nu}{2\pi}\int A\wedge (d\omega^0+\omega_1\wedge \omega_2+\frac{1}{\ell^2} e_1 \wedge e_2)\,,
\end{align}
forcing the  mean orbital angular momentum $\bar{s}=1$.
%average effective angular momentum carried by the electrons in the cyclotron motion as $\bar{s}=1/2$.
We point out that $\omega^0_\mu$ is equal to $\widehat{\omega}_{\mu }$. All these reduce the action to
\begin{align}
S_{\mathfrak{C}_4}[e,\omega , A]=&\frac{\nu }{4\pi}\int A d A +\frac{\nu}{2\pi}\int A\wedge (d\omega^0+\omega_1\wedge \omega_2+\frac{1}{\ell^2} e_1 \wedge e_2)\nonumber\\
&+\frac{\nu}{4\pi}\frac{ \alpha_0}{ \alpha_2}\,\int CS(\omega )+\frac{\nu}{4\pi}\frac{1}{\ell^2}\, \int e_a \wedge D_{\omega }e^a\,.\nonumber
\end{align}
Note that the torsion is now fueled by $D_\omega e_a=-e_b\wedge A\,\epsilon_a^{~b0}$, which means contribution to the normally torsionless connection $\omega$ used in the Wen-Zee term \cite{Geracie:2014mta}.

The stress-energy tensor is obtained by the variation of the dreibein
\begin{align}
\mathfrak{T}^\mu_a=\frac{\delta S_{\mathfrak{C}_4}}{\delta e^a_\mu}=\frac{\nu}{4\pi}\frac{2}{\ell^2}\,\epsilon^{\mu\rho\sigma} (\frac{1}{2} T_{a\,\rho\sigma}+e_{b\,\rho}\, A_\sigma\,\epsilon_a^{~b0})\,.
\end{align}
The Hall viscosity coefficient, standing in front of the torsional term and describing the response of the quantum Hall fluid, is defined as $\eta_H=\frac{\nu\bar{s}B_0}{4\pi}$ \cite{Hughes:2012vg,Geracie:2014mta}. This is in agreement with our framework after relating $\ell$ to the magnetic length $\ell=\sqrt{\hbar c/|e|B_0}$ with $B_0$ being an external magnetic field \cite{Cappelli:2015ocj}. Then we have  $\frac{\nu}{4\pi}\frac{1}{\ell^2}=\frac{\nu B_0}{4\pi}=\eta_H$ with once again $\bar{s}=1$.
%\begin{align}
%\mathfrak{T}^\mu_a=\frac{\delta S_{\mathfrak{C}_4}}{\delta e^a_\mu}=\frac{\nu}{4\pi}\frac{2}{\ell^2}(D_\omega e_a+e_b\wedge A\,\epsilon_a^{~b0})
%\end{align}

Lastly, the value of the constant $\alpha_0$ in front of the CS$(\omega)$ term can be a subject of considerations related to framing anomaly of the Chern-Simons theory and the redefinition related to the central charge of a conformal boundary \cite{Gromov:2014dfa}. Notice, however, that the field equations in the case of $\alpha_0\neq\alpha_2$ offer splitting into $R^0=0$ and $(dA+\frac{1}{2\ell^2}\epsilon^{0bc}\,e_b\wedge e_c)=0$, whereas $\alpha_0=\alpha_2$ leads to $R^0=-(dA+\frac{1}{2\ell^2}\epsilon^{0bc}\,e_b\wedge e_c)$.

To conclude, we see from the analysis of the six gauge algebras that it is not the Maxwell algebra $\mathfrak{B}_4$ (which misses $\frac{1}{\ell^2}\epsilon^{abc}\hat{A}_a e_b e_c$ and more importantly $\hat{A}_a\wedge D_{\omega }\hat{A}^a$ terms) but the  $\mathfrak{C}_4$ generalization of it that gives the widest scope of the topological insulators framework. Not only it makes the straightforward construction possible, but it also provides a concise set of constants. Different algebras deliver limited setups but they might be useful in some contexts, for instance, only $\mathfrak{C}_4$ has a non-vanishing torsion. Configurations like $B_4,\tilde{B}_4$, and $\mathfrak{B}_4$ do not have $AdA$ term (therefore we are without the normal Hall '$E$' and '$B$' current responses), while, interestingly, they still do have the Wen-Zee term present.

\section{Conclusions}

In the recent years, we have observed a growing number of works exploring the framework and geometrical methods originally introduced in the context of the General Relativity, being now applied to the various phenomena in the field of condensed matter physics. In this paper, we dealt with the construction of the effective action that exploits the notion of the Maxwell/Poincar\'e/(A)dS algebra enlargements carried by the Chern-Simons model. This seems to provide interesting structure capturing crucial elements of the description of the topological insulator. The resonant algebras \cite{Durka:2016eun} resulting from the semigroup expansion (S-expansion) were a very convenient tool to carry the whole action construction extending the results of \cite{Palumbo:2016nku}. 

\section{Acknowledgment}

We would like to thank prof.\ Jerzy Lukierski for bringing to our attention the publication \cite{Palumbo:2016nku}, which was the starting point of this article. For RD this work is supported by Institute Grant for Young Researchers 0420/2716/18 and for JKG  this work is supported by funds provided by the National Science Center, projects number 2017/27/B/ST2/01902.

\end{document}